\documentclass[aps,pre,twocolumn]{revtex4-2}
\usepackage{amssymb}
\usepackage{amsmath,bm}
\usepackage{graphicx,color}
\usepackage{epsfig}
\usepackage{multirow}
\usepackage[cal=boondox]{mathalpha}
\usepackage{accents}
\DeclareMathAccent{\wtilde}{\mathord}{largesymbols}{"65}
\usepackage{stackengine}
\stackMath
\newcommand\tenq[2][1]{%
	\def\useanchorwidth{T}%
	\ifnum#1>1%
	\stackunder[0pt]{\tenq[\numexpr#1-1\relax]{#2}}{\scriptscriptstyle\sim}%
	\else%
	\stackunder[1pt]{#2}{\scriptscriptstyle\sim}%
	\fi%
}

\begin{document}

\title{Experimental Confirmation of Hyperuniformity of torque fluctuations in Frictional Matter}

\author{Jin Shang$^{1,2}$ and Jie Zhang$^{2,3}$ Itamar Procaccia$^{1,4}$}
\affiliation{$^1$Hangzhou International Innovation Institute, Beihang University, Hangzhou, China, 311115\\$^2$School of Physics and Astronomy, Shanghai Jiao Tong University, 800 Dong Chuan Road, 200240 Shanghai, China $^3$Institute of Natural Sciences, Shanghai Jiao Tong University, 200240 Shanghai, China,$^4$Department of Chemical Physics, the Weizmann Institute of Science, Rehovot 7610001, Israel.}

\begin{abstract}
A recent purely theoretical prediction stated that in frictional granular packings, mechanical balance and material isotropy constrain the stress auto-correlation matrix to be fully determined by two spatially isotropic functions: the pressure
and torque auto-correlations. Moreover, and unexpectedly, the torque fluctuations were predicted to be hyperuniform, a condition for the stress auto-correlation to decay as the elastic Green's function. In this Letter we present first experimental evidence for the hyperuniformity of the stress fluctuations. We propose that quite generally the variance of torque fluctuations in a d-dimensional ball of radius R will increase exactly like $R^{d-1}$, satisfying the definition of hyperuniformity.  
\end{abstract}

\maketitle

{\bf Introduction:}  The concept of hyperuniformity appears in many scientific contexts, from Solid State Physics \cite{03TS}, through Biology \cite{14JLHMCT}, all the way to Cosmology \cite{02GJS}. The term hyperuniformity was coined by Torquato and Stillinger in Ref.~\cite{03TS}. In cosmological contexts the concept was  independently called super-homogeneity \cite{02GJS}. In essence, hyperuniformity refers to the nature of the fluctuations of a density, be it a number density, a vector or tensor density, at large scales.  Denoting the density as $q$, and the integral of $q$ over a d-dimensional ball of radius R as $Q$. The issue is how the variance of $Q$ grows upon increasing the radius $R$. ``Normal" fluctuations exhibit variance that grows like $R^d$ in $d$-dimensions. When the variance grows slower than $R^d$ for large $R$, the density if referred to as hyperuniform. Precisely, a density $q$ is hyperuniform when  
\begin{equation}
\lim_{R\to \infty} \frac{	\langle (Q-\langle Q\rangle)^2\rangle }{R^d} =0 \ . 
\end{equation}
Here $\langle \cdots\rangle$ denotes an average over realizations.  A consequence of this property is the vanishing of the structure factor at the origin:
\begin{equation}
	\lim_ {k \to  0} S ( \mathbf{k} ) = 0 \quad \text {for wave vectors} ~\mathbf{k} \in R^d  .
\end{equation}

While useful in many context, hyperuniformity appeared unexpectedly in a recent theoretical paper dealing with frictional amorphous solids. By ``frictional" one means assemblies of particles that in addition to having normal forces exerted on each other, they have also tangential frictional forces, losing the Hamiltonian symmetry that characterizes frictionless particles. Needless to say, such materials are everywhere, from sand piles to powders, from apples on the market stall to the gouge in a geological fault. The theoretical prediction is that in a very broad variety of such systems the torque density is hyperuniform. Moreover, this property is a necessary condition for the stress correlations in such system to decay as predicted by elasticity theory at large distances. The aim of this Letter is to present the first experimental verification of this prediction.  

{\bf Theoretical Background:} During the last decade it became clear that the stress field of amorphous solids whose inter-particle forces derive from a Hamiltonian, present long ranged correlation tails of a form similar to elastic Green's functions \cite{09HC,14Lem,15Lem,17Lem,18Lem}. It was demonstrated that these long range correlations follow in Hamiltonian problems from the conjunction of three properties. These are (i) Mechanical balance, (ii) Material isotropy and (iii) the normality of local pressure fluctuations \cite{17Lem,18Lem}. The derivation of these results depends crucially on the symmetry of local stress which inevitably breaks down in the presence of frictional forces which introduce local torques. The question was then fully open about the nature of stress correlations in frictional granular packings, an important, diverse and widespread class of materials as mentioned above.

The question was answered in a recent theoretical paper \cite{21LMPR}. It was shown again that  in Hamiltonian systems with central forces, mechanical balance and material isotropy demand the stress auto-correlation matrix to be fully determined by the pressure auto-correlation only. In sharp contrast, it was shown that in  frictional granular packings, it is determined not by one but by two spatially isotropic functions, the pressure and torque auto-correlations. It was demonstrated that in the absence of external
torques, the torque fluctuations are hyperuniform, i.e. the torque auto-correlation vanishes in the
zero wave-number limit. As a consequence the torque contribution to the stress auto-correlation is
sub-dominant at large wave-length. Consequently, the large distance decay of the stress auto-correlation is again determined by the scaling of local pressure fluctuations on domains of increasing sizes. When these fluctuations are normal the presence of elastic-like long-ranged  contributions follows. The relevance of these theoretical predictions to a wide array of frictional materials warrants an experimental verification, and this is the aim of this Letter.

\begin{figure}
    \centering
	\includegraphics[width= 8.6 cm]{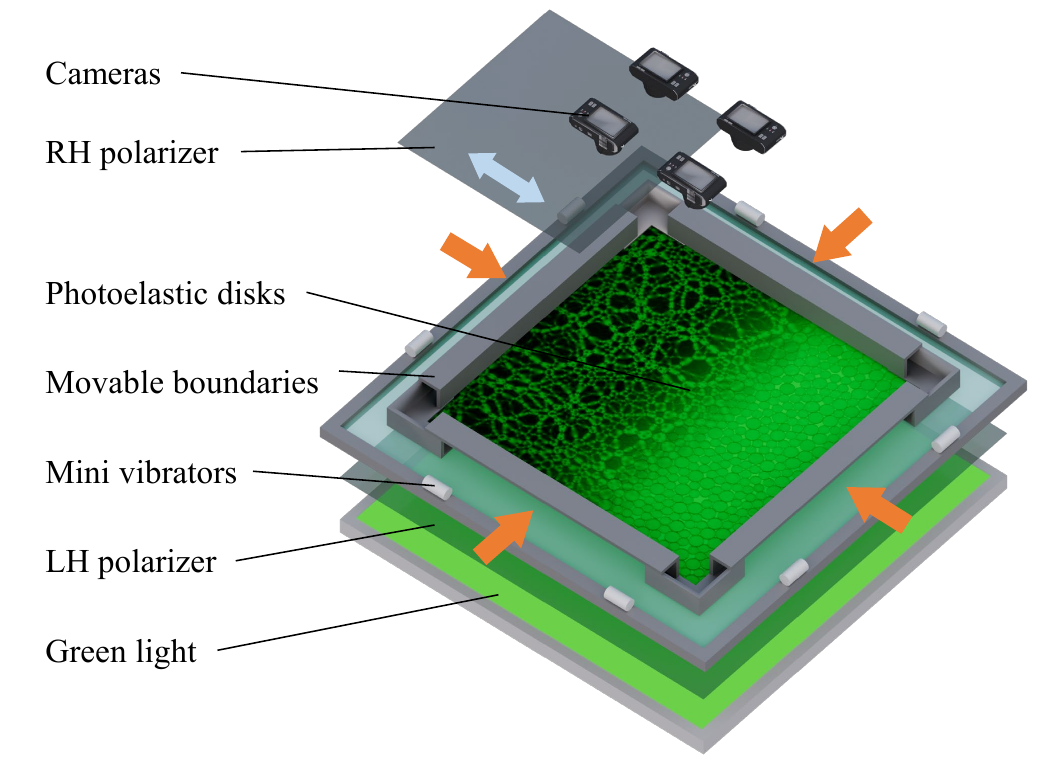}
	\caption{Schematic of the experimental setup. Inside we show a typical image in which half the cell exhibits disks and the other half shows force chains.}
	\label{apparat}
\end{figure}

{\bf Materials and Methods:} The experimental apparatus, cf. Fig.~\ref{apparat}, is composed of a biaxially movable square frame mounted on a horizontal glass substrate. The apparatus is filled with a granular packing comprising 2710 small and 1350 large photoelastic disks (in fact, flat cylinders), made of Vishay PSM-4, with diameters of 10 mm and 14 mm, respectively. The sidewalls of the square frame are capable of inward movement at a speed of 0.1 mm/s, thereby imposing quasi-static isotropic compression on the inner granular packing. 
To minimize the influence of friction between the disks and the underlying glass substrate, the glass surface is coated with a thin layer of talcum powder. Additionally, mini vibration motors, affixed to the aluminum frame supporting the glass plate, are activated during boundary motion. 

A green LED light source is positioned beneath the apparatus, overlaid with a left-handed circular polarizer. Particle configurations and stress distributions are captured via a $2\times2$ array of high-resolution (10 pixels/mm) digital cameras, located about 1.5 m above the apparatus.  A movable right-handed circular polarizer is installed in front of the cameras. When the polarizer is removed, normal images are captured to determine the spatial positions of the disks. Upon insertion of the polarizer, photoelastic stress images are obtained, and a force-inversion algorithm \cite{05MB,Wang2020nc,Wang2021prr} is used to reconstruct the vector contact forces between the disks.

To generate a jammed packing at a specific pressure, an initial stress-free, random, and homogeneous configuration is prepared at a packing fraction $\phi = 0.834$ slightly below the jamming point $\phi_J\approx 0.84$ of the corresponding frictionless configuration\cite{OHern2002prl}. Subsequently, quasi-static isotropic compression is applied step by step until the target packing fraction $\phi = 0.862$ is reached, corresponding to an average pressure of $p = 31.1 \ \mathrm{N\,m^{-1}}$. This preparation protocol yields force networks that are isotropic and approximately homogeneous. In the description of the experimental results below we choose the radius of the small disk as the unit of length. To mitigate potential boundary effects, statistical analyses exclude about 10 layers of disks near the boundaries.

\begin{figure}[!htb]
    \centering
	\includegraphics[width= 8.6cm]{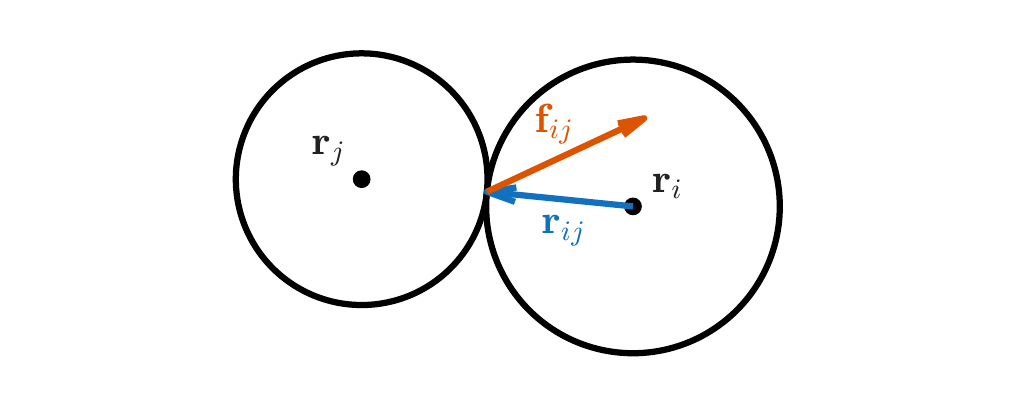}
	\caption{Definitions of $\mathbf{r}_i$, $\mathbf{r}_{ij}$ and $\mathbf{f}_{ij}$.}
	\label{definition}
\end{figure}

The torque and pressure are defined at each contact, without any coarse graining. The torque and pressure fields can be expressed as
\begin{equation}
  \tau(\mathbf{r}) = -\sum_{i,j}(\mathbf{r}_{ij} \times \mathbf{f}_{ij}) \cdot \hat{\mathbf{z}} \delta [\mathbf{r} - (\mathbf{r}_i + \mathbf{r}_{ij})],
\end{equation}
\begin{equation}
  p(\mathbf{r}) = -\sum_{i,j} \mathbf{r}_{ij} \cdot \mathbf{f}_{ij} \delta [\mathbf{r} - (\mathbf{r}_i + \mathbf{r}_{ij})],
\end{equation}
where the definitions of the variables are shown in the Fig.~\ref{definition}. The unit vector  $\hat{\mathbf{z}}$ is orthogonal to the plane of the disks.

{\bf Experimental Results}: To assess the nature of the fluctuations of our two fundamental fields,  we integrate the torque (or pressure) within a circular window of a random center $\mathbf{r}$ and a radius $R$. The sampling is repeated 4000 times for each window radius and then the variance of these integrals are calculated. For torque, it is
\begin{equation}
  \sigma_{\tau}^2(R) = \mathrm{Var} \left ( \sum_{|\mathbf{r}_{i} + \mathbf{r}_{ij} - \mathbf{r} |<R} (\mathbf{r}_{ij} \times \mathbf{f}_{ij}) \cdot \hat{\mathbf{z}} \right ),
\end{equation}
and similarly for the pressure. 
\begin{figure*}[!htb]
    \centering
    \includegraphics[width= 17.8 cm]{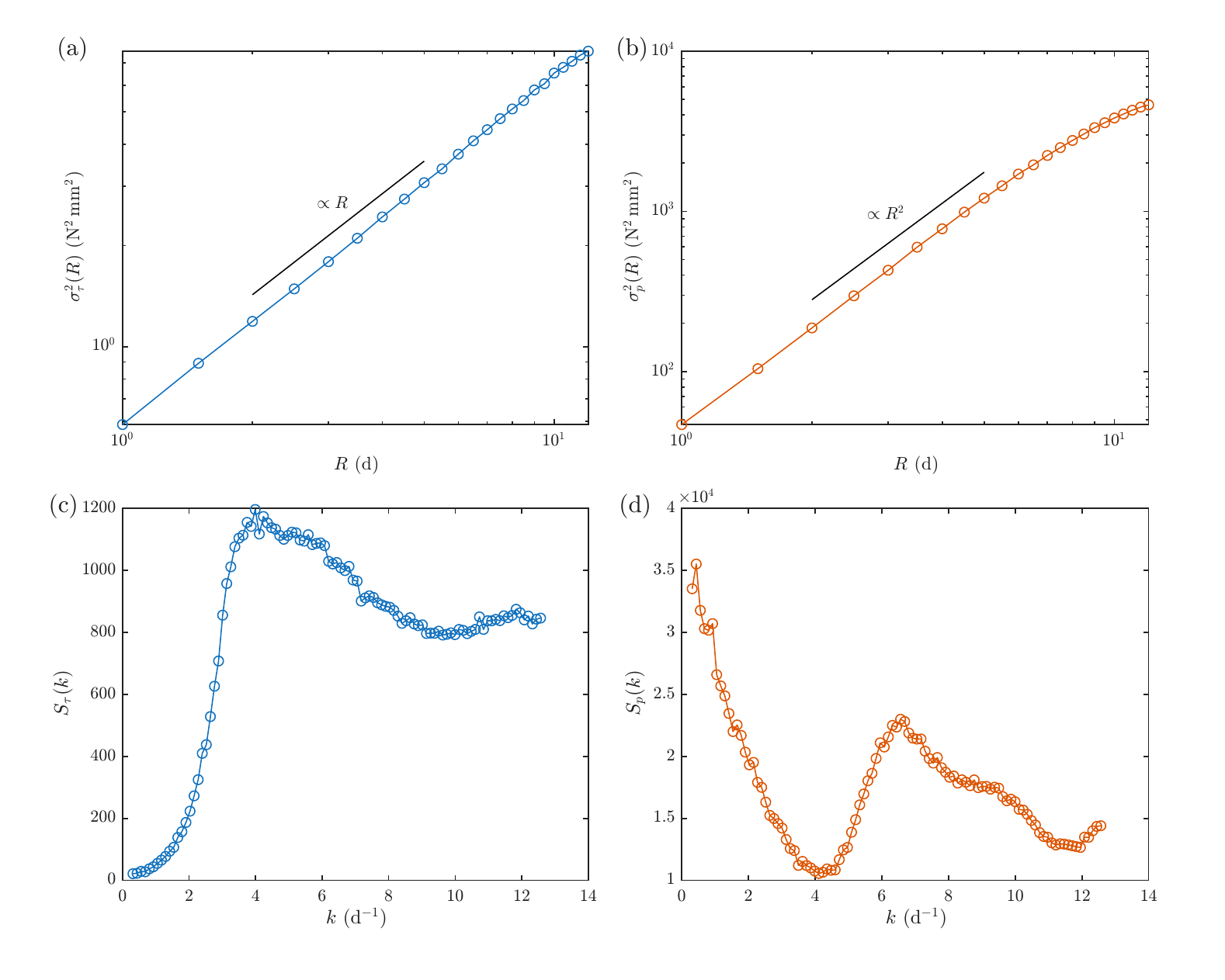}
    \caption{(a) The $R$ dependence of the variance of of torque fluctuations. The linearity in R is a direct evidence of hyperuniformity. Below we explain that the result is exact, i.e. the growth of variance of the torque is like $R^\alpha$ with $\alpha=1$. (b) The $R$ dependence of the variance of of pressure fluctuations. (c) $S(k)$ of torque. (d) $S(k)$ of pressure.}
    \label{FluctSK}
\end{figure*}


As shown in the Fig.~\ref{FluctSK} (a) and (b), the variance of the torque fluctuation increases like $R$ and for pressure like $R^2$. The deviations at large scales are due to the finite-size effect. 

Finally, we examine the Fourier spectra of our field fluctuations.  
The spectrum of the torque fluctuations is defined (with a similar definition for the pressure) as
\begin{equation}
  S_{\tau}(\mathbf{k}) = \left | \sum_{i,j}\delta \tau_{ij} \mathrm{e}^{-\mathrm{i} \mathbf{k} \cdot (\mathbf{r}_i + \mathbf{r}_{ij})} \right |^2
\end{equation}
with $\tau_{ij} = (\mathbf{r}_{ij} \times \mathbf{f}_{ij}) \cdot \hat{\mathbf{z}}$ and $\delta \tau_{ij}  = \tau_{ij} - \langle \tau_{ij} \rangle$.
After an angular averaging each spectrum on $\mathbf{k}$, and then averaging over 30 independent samples, the results are shown in Fig.~\ref{FluctSK} (c) and (d). The vanishing of the spectrum of the torque fluctuations at $k\to 0$ is a direct evidence of hyperuniformity. 

\begin{figure}
	\centering
	\includegraphics[width= 8.6cm]{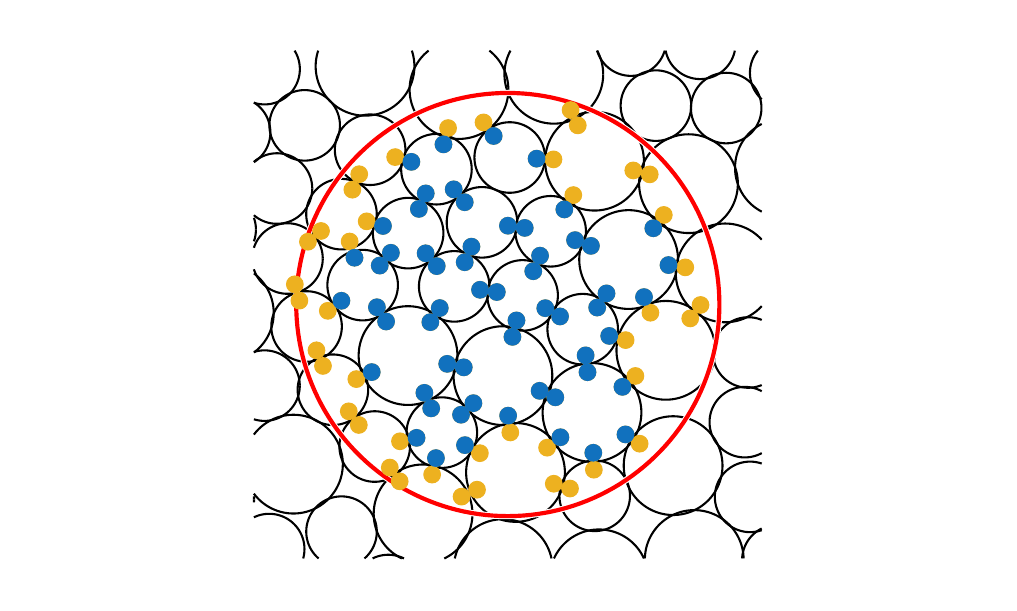}
	\caption{Illustrating the consequences of the constraint of the vanishing of the torque for each particle. See text for explanation}
	\label{constraint}
\end{figure}

While the normal fluctuations of the pressure with a variance growing like $R^2$ is expected and observed, the torque variance growing like $R$ calls for an explanation. The reason for this is quite intuitive, and results from the constraint on the resultant torque which must vanish on each particle. The consequences of this constraint can be seen in Fig. ~\ref{constraint}. Integrating the torque inside any randomly chosen circle (outer red line), one gets no contribution from the inner contacts (highlighted as pairs of blue points), since they are totally within the circle, they contribute exactly zero. The only contribution comes form contacts that are adjacent to the outer boundary, where some contacts are within (highlighted  as pairs of yellow points) and some are outside the circle (not highlighted ). Since the circle cuts the configuration randomly, the values of torque computed along the boundary are random, and will exhibit a normal variance increase proportional to the length of the boundary, in agreement with the observation presented in Fig.~\ref{FluctSK} (a). 

{\bf Discussion}: without deeper scrutiny, one could suspect that sand, or powders, or any other common frictional assembly of particles, may not exhibit standard ``elastic like" decay of stress correlations at large distances. Contrary to systems with Hamiltonian interaction, where the crucial property, on top of isotropy and mechanical balance, is normal fluctuations of the pressure, in frictional matter a second condition is necessary, the hyperuniformity of the torque fluctuations. These two conditions guarantee that even sand or powders should appears ``elastic" at large scale from the point of view of stress decay, as measured quantitatively in the 2D experiment\cite{Wang2020nc}. While this was made theoretically clear in Ref.~\cite{21LMPR}, the results presented above offer the first experimental verification that this is indeed the nature of pressure and torque fluctuations in frictional matter. The hyperuniformity of the stress fluctuations guarantees that at large scale the pressure fluctuation dominate, leading to a situation that is analogous to Hamiltonian systems in which normal fluctuation of pressure are necessary and sufficient for elastic like behavior, irrespective of the violent breaking of Hamiltonian symmetry. 

At this point it is worthwhile to comment that the result pertaining to the growth of the variance of the torque fluctuations with a power law $R^{d-1}$ in $d$ dimensions is very generic. In other words, even if we employ non spherical frictional particles, in which the torques will have contributions that are not purely frictional, the constraint on the vanishing of the resultant torque on each particles will lead, as explained in Fig.~\ref{constraint}, to the same conclusion, i.e. a power law of the form $R^{d-1}$. In a future follow-up paper we will present additional experimental results for non-spherical particles and systems under shear to support these generic expectations. 

The present experimental set-up that allows measurement of the force vector at every contact is very useful for the investigations reported above, but also rather uncommon. The apparent generality of the theoretical condition of hyperuniformity of the torque fluctuations for elastic behavior needs further experimental verification. It is our hope that other methods to study this fundamental issue in other assemblies of frictional particles will follow suit. 

{\bf Acknowledgments.} JS and JZ thank Yinqiao Wang for helpful discussions. JS and JZ acknowledge the support of the NSFC (Nos. 11974238 and 12274291) and also the support of the Innovation Program of Shanghai Municipal Education Commission under No. 2021-01-07-00-02-E00138. JS and JZ also acknowledge the support from the Student Innovation Center of Shanghai Jiao Tong University.

\bibliography{All}

\end{document}